\documentclass[prl,twocolumn]{revtex4}
\usepackage{graphicx}
\usepackage{amssymb}
\usepackage{color}

\begin{document}
\title{Controlling a magnetic Feshbach resonance with laser light}
\author{Dominik M. Bauer}
\author{Matthias Lettner}
\author{Christoph Vo}
\author{Gerhard Rempe}
\author{Stephan D\"{u}rr\footnote{e-mail: stephan.duerr@mpq.mpg.de}}
\affiliation{Max-Planck-Institut f{\"u}r Quantenoptik, Hans-Kopfermann-Stra{\ss}e 1, 85748 Garching, Germany}

\hyphenation{Feshbach}

\maketitle

{\bf 
The capability to tune the strength of the elastic interparticle interaction is crucial for many experiments with ultracold gases. Magnetic Feshbach resonances \cite{tiesinga:93,inouye:98,chin:0812.1496} are a tool widely used for this purpose, but future experiments \cite{garay:00,carusotto:08,rodas-verde:05,dong:06,abdullaev:08,deng:08} would benefit from additional flexibility such as spatial modulation of the interaction strength on short length scales. Optical Feshbach resonances \cite{fedichev:96a,bohn:97,fatemi:00,theis:04,thalhammer:05,jones:06} offer this possibility in principle, but suffer from fast particle loss due to light-induced inelastic collisions. Here we show that light near-resonant with a molecular bound-to-bound transition can be used to shift the magnetic field at which a magnetic Feshbach resonance occurs. This makes it possible to tune the interaction strength with laser light and at the same time induce considerably less loss than an optical Feshbach resonance would do.
}

Using light to change the $s$-wave scattering length $a$ in ultracold gases offers more flexibility than a magnetic Feshbach resonance (FR) because it is possible to apply an almost arbitrary spatial pattern of light using holographic masks. The light intensity can vary on a length scale of typically one optical wavelength and the pattern can additionally be varied rapidly in time. This could be used for a variety of applications, such as the simulation of the physics of black holes \cite{garay:00,carusotto:08}, the controlled creation of solitons \cite{rodas-verde:05}, studies of the collapse of a BEC in an unusual regime \cite{dong:06}, and the simulation of certain Hamiltonians, in which the scattering length needs to be different at different sites of an optical lattice \cite{abdullaev:08,deng:08}. Moreover, if each lattice site contains exactly two atoms \cite{volz:06} and $a$ is varied only on every second lattice site, one could associate molecules at every second lattice site by ramping the magnetic field across the FR, thus producing a quantum state which resembles that of a supersolid. Another possible application for the manipulation of $a$ with light exists in gases consisting of a mixture of different species or spin states. It would be desirable to tune the various scattering lengths in such systems independently, but for that purpose more control parameters than just the magnetic field are needed. Furthermore, if a spatially random light intensity pattern is applied, the scattering length would vary randomly with position which might give rise to new quantum phases of the atomic gas.

A known scheme to manipulate $a$ using light employs a photoassociation (PA) resonance, sometimes also called optical Feshbach resonance. But so far, PA resonances have rarely been used to tune $a$ because they induce fast particle loss. The experiments in Refs.\ \cite{theis:04,thalhammer:05} both demonstrated a change of ${\rm Re}(a)/a_{\rm bg}-1 \sim \pm1$ in $^{87}$Rb, where $a_{\rm bg}$ is the background value of $a$. For these parameters, both experiments incurred losses characterised by a two-body rate coefficient $K_2$ with an estimated value of $\sim 10^{-10}$ cm$^3$/s. Typical densities on the order of $10^{14}$ cm$^{-3}$ result in lifetimes on the order of 100 $\mu$s, which is too short for many applications.

In this Letter we experimentally demonstrate that laser light can noticeably shift a magnetic FR and at the same time induce considerably smaller particle loss rates than an optical Feshbach resonance. We study the magnetic field dependence of ${\rm Re}(a)$ for a large detuning of the laser from a bound-to-bound resonance and find a change in ${\rm Re}(a)$ similar to that reported for optical Feshbach resonances. In addition, we study the system for smaller detuning and observe an Autler-Townes doublet in the particle loss. Our results are in good agreement with a theoretical model that we develop.

\begin{figure}[t!]
\includegraphics[width=0.7\columnwidth]{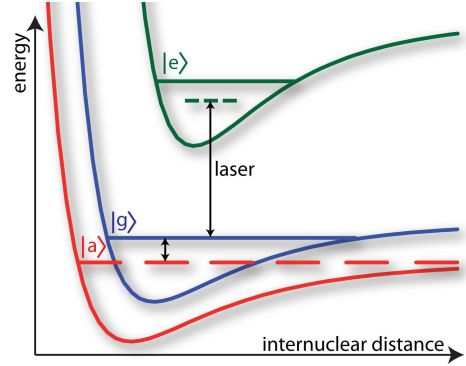}
\caption{
\label{fig-level-scheme}
Level scheme of the experiment. The Feshbach resonance couples atoms at threshold in the incoming channel $|a\rangle$ to a bound dimer state $|g\rangle$ in a different potential. A laser is near resonant with a bound-to-bound transition from $|g\rangle$ to an electronically excited dimer state $|e\rangle$.}
\end{figure}

A basic level scheme for our experiment is shown in Fig.\ \ref{fig-level-scheme}. A light field is near-resonant with a bound-to-bound transition from the dimer state $|g\rangle$ in the electronic ground state to an electronically excited dimer state $|e\rangle$. The FR coupling between state $|g\rangle$ and the free atom state $|a\rangle$ is typically much weaker than the bound-to-bound coupling due to the light field. In order to understand the physics, it is therefore useful to consider the energy eigenstates that are created from the two dimer states when applying the light. These eigenstates are superpositions of states $|g\rangle$ and $|e\rangle$ with corresponding energy shifts. The FR coupling probes the $|g\rangle$ components of both energy eigenstates, resulting in two resonances as a function of magnetic field $B$, in analogy to an Autler-Townes doublet \cite{autler:55,cohen-tannoudji:92}. Population in state $|e\rangle$ undergoes spontaneous radiative decay predominantly into other levels, not shown in Fig.\ \ref{fig-level-scheme}. This leads to loss of population from the system. With all the population initially in state $|a\rangle$ this loss can be described by a two-body loss rate coefficient $K_2$, as discussed in the Methods section.

\begin{figure}[t!]
\includegraphics[width=0.9\columnwidth]{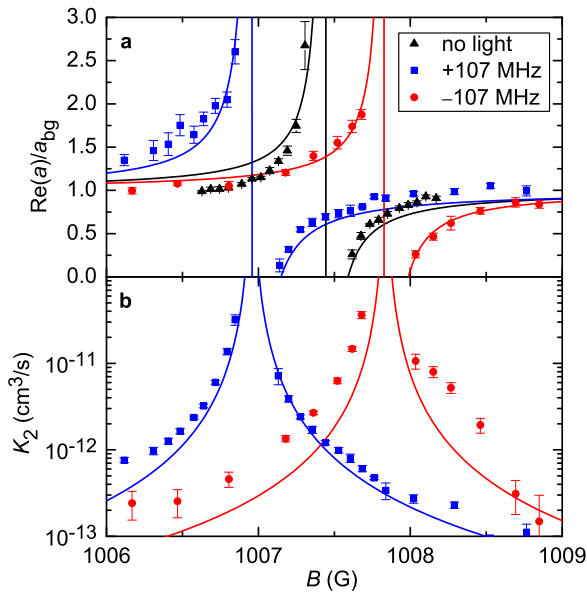}
\caption{
\label{fig-a-vs-B}
Shifting a Feshbach resonance with laser light. {\bf a} Real part of the scattering length as a function of magnetic field. In the absence of light ($\blacktriangle$), the pole in the scattering length occurs a 1007.4 G. With 4.2 mW of light applied the Feshbach resonance is shifted to a different magnetic field. The data were recorded at a detuning of the light frequency with respect to the bound-to-bound resonance of $\Delta_L/2\pi=+107$ MHz
({\color{blue}$\scriptstyle \blacksquare$}) and $-107$ MHz ({\color{red}$\bullet$}). {\bf b} The application of the light induces two-body loss described by the rate coefficient $K_2$ for a BEC. The observed loss when changing ${\rm Re}(a)/a_{\rm bg}$ by $\pm1$ is typically one order of magnitude slower than the loss that would be incurred when using an optical Feshbach resonance. Error bars represent one statistical standard error.}
\end{figure}

The experimental procedure starts with a Bose-Ein\-stein condensate (BEC) of $^{87}$Rb atoms in the hyperfine state $|F=1,m_F=1\rangle$ in an optical dipole trap with trap frequencies $(\omega_x,\omega_y,\omega_z)/2\pi=(74,33,33)$ Hz. The magnetic field $B$ is oriented along the $z$ axis and held at a value that is a few gauss away from the FR at 1007.4 G \cite{marte:02}. At time $t=0$, $B$ is jumped to a certain value at which it is held for 2 ms. The dipole trap is typically also switched off at $t=0$, except for Fig.\ \ref{fig-a-vs-B} where it is turned off at $t=2$ ms. During the 2 ms hold time of $B$, a travelling-wave laser beam is applied. This beam propagates along the vertical $x$ direction, has a waist ($1/e^2$ radius of intensity) of 0.17 mm, and is $\pi$ polarised, {\it i.e.}, the electric field oscillates along the $z$ axis. The laser is beat locked to a precision frequency comb and drives the bound-to-bound transition. After the 2 ms hold time of $B$, the laser and $B$ are switched off simultaneously. This is followed by free flight and the atom cloud is imaged 18 ms after release from the dipole trap. The evolution of the cloud size during this sequence can be modelled in analogy to Refs.\ \cite{castin:96,volz:03}. In addition, we include two-body loss in the corresponding equations of motion. This model is used to extract $K_2$ and ${\rm Re}(a)$ from the measured atom number and cloud size after expansion.

The experimental results in Fig.\ \ref{fig-a-vs-B}{\bf a} clearly show that the magnetic Feshbach resonance is shifted by $\sim \pm 0.5$ G when applying the laser light. The light is far detuned from the bound-to-bound transition. This causes an ac-Stark shift of state $|g\rangle$, which results in a corresponding shift of the magnetic field at which the FR occurs.

Fig.\ \ref{fig-a-vs-B}{\bf b} shows the corresponding two-body loss rate coefficient $K_2$ for a BEC. The solid lines are the predictions for $K_2$ based on the model and parameter values presented in the Methods section. In the absence of light, $K_2$ is zero. With the light on, we attribute all observed particle loss to $K_2$. This overestimates $K_2$ on the low-field side of the resonance, where part of the loss is actually caused by inelastic three-body collisions \cite{smirne:07}. On the high-field side of the resonance, the three-body loss is less important and the extracted values for $K_2$ for blue detuning ({\color{blue}$\scriptstyle \blacksquare$}) agree well with the solid line. The agreement for red detuning ({\color{red}$\bullet$}) is not quite as good.

In order to compare with an optical Feshbach resonance, we consider the data points in Fig.\ \ref{fig-a-vs-B}{\bf a} with ${\rm Re}(a)/a_{\rm bg}-1\sim \pm1$. For these data points Fig.\ \ref{fig-a-vs-B}{\bf b} shows $K_2\sim 10^{-11}$ cm$^3$/s, which is an improvement by one order of magnitude compared to Refs.\ \cite{theis:04,thalhammer:05}.

Our technique offers two scenarios for applications. In the first scenario, the magnetic field can be held so close to the position of the unshifted FR, that this already changes $a$ substantially. The light can then be used to shift the position of the FR in magnetic field, thus bringing $a$ back to $a_{\rm bg}$ in regions of high light intensity, while regions of low light intensity remain at a value far away from $a_{\rm bg}$. In the second scenario, the magnetic field is held further away from the unshifted FR so that $a\sim a_{\rm bg}$. Application of the light can shift the FR so close to the chosen value of $B$ that $a$ is modified strongly in regions of high light intensity. In both scenarios, the light only causes a moderate shift of the FR, it does not need to induce a direct PA coupling between the atomic and molecular state, which would result in loss just like for an optical Feshbach resonance.

\begin{figure*}[t!]
\includegraphics[scale=1]{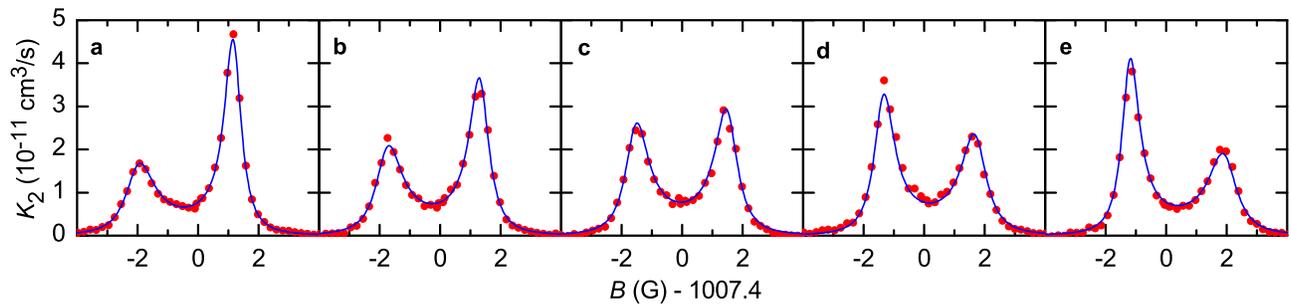}
\caption{
\label{fig-Autler-Townes}
Autler-Townes splitting of a magnetic Feshbach resonance caused by application of a laser that drives a bound-to-bound transition. The loss rate coefficient $K_2$ is measured as a function of magnetic field $B$ for different values of the laser frequency, that increase from {\bf a} to {\bf e} in steps of 1 MHz. Part {\bf c} is recorded at 382,046,942.62 MHz very close to resonance, where the Autler-Townes doublet becomes symmetric. The lines show a simultaneous fit to all these data sets.}
\end{figure*}

To further explore the physics of our system we study the regime of much smaller detuning of the laser light. In this regime we clearly observe an Autler-Townes doublet as a function of $B$, as shown in Fig.\ \ref{fig-Autler-Townes} for a laser power of $P=0.47$ mW. We fit the model from the Methods section to all the data in this figure simultaneously. The fit agrees well with the experimental data.

\begin{figure}[t!]
\includegraphics[width=0.9\columnwidth]{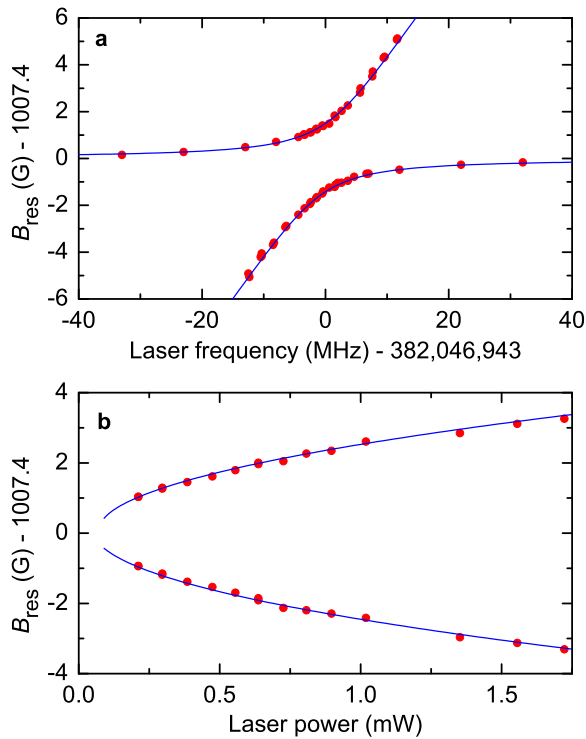}
\caption{
\label{fig-avoided-crossing}
Systematic study of the resonance position. The magnetic fields at which loss resonances occur are shown as a function of laser frequency ({\bf a}) and laser power ({\bf b}). Part {\bf a} was recorded at a power of 0.47 mW and clearly shows an avoided level crossing. Part {\bf b} was recorded at a laser frequency of 382,046,943 MHz. The solid lines show the prediction of the model in the Methods section with the parameter values obtained from Fig.\ \ref{fig-Autler-Townes}.}
\end{figure}

We also recorded the position of the two loss resonances systematically as a function of laser frequency and laser power, as shown in Fig.\ \ref{fig-avoided-crossing}. The data exhibit typical features of an Autler-Townes doublet, namely an avoided level crossing in {\bf a} and a splitting approximately proportional to the square root of the laser power in {\bf b}. The data are well described by the model in the Methods section. We conclude that the relevant physics is well understood.

The technique described here could be improved even further by using larger detuning combined with higher laser power, which should reduce the loss. Alternatively, the loss could be reduced by replacing the electronically excited dimer state $|e\rangle$ by a dimer state in the electronic ground state. The coupling to this state could be achieved with a stimulated two-photon Raman transition. Some potential applications of our method were already mentioned. A further example is to shift one FR resonance on top of another to explore the coupling between the two. Yet another possible application can be found in $^{133}$Cs. Its hyperfine state $|F=3,m_F=3\rangle$ has a very broad FR at a slightly negative magnetic field. Applications of this FR are hampered by fast loss due to dipolar relaxation at negative magnetic field. Shifting this FR to a positive magnetic field would make it accessible for experiments, such as studies of Efimov physics \cite{kraemer:06}.

\section*{Methods}

\subsection*{Theoretical Model}
A theoretical model for the experiment can be developed as follows. We denote the atomic state as $|a\rangle$, the ground state dimer as $|g\rangle$, and the electronically excited dimer state as $|e\rangle$. We describe the populations of the states with the mean fields $\psi_a$, $\psi_g$, and $\psi_e$. Generalising Ref.\ \cite{timmermans:99} we use the following equations
\begin{eqnarray}
i \frac{d}{dt} \psi_a &=& \frac{U_{\rm bg}}{\hbar} |\psi_a|^2 \psi_a + 2 \alpha^* \psi_a^* \psi_g + 2 \beta^* \psi_a^* \psi_e \\
i \frac{d}{dt} \psi_g &=& \alpha \psi_a^2 + \Delta_g \psi_g + \frac12 \Omega_R^* \psi_e \\
i \frac{d}{dt} \psi_e &=&  \beta \psi_a^2 + \frac12 \Omega_R \psi_g + \Delta_e \psi_e - \frac i2 \gamma_e \psi_e
.
\end{eqnarray}
$U_{\rm bg} = 4\pi\hbar^2 a_{\rm bg}/m$ describes the mean-field energy far away from the FR, where $m$ is the mass of one atom and $\hbar$ is the reduced Planck constant. The electric field of the light is $E=-E_0 \cos(\omega_Lt)$ with amplitude $E_0$ and angular frequency $\omega_L$. This field causes a coupling on the bound-to-bound transition $|g\rangle \leftrightarrow |e\rangle$ with a resonant Rabi frequency $\Omega_R=d_{eg} E_0/\hbar$, where $d_{eg}=\langle e|d|g\rangle$ is the matrix element of the electric dipole moment. $\gamma_e$ describes spontaneous radiative decay of state $|e\rangle$ into other states, which are not included in the model. Collisional decay is neglected.

The energy of each state depends nonlinearly on the magnetic field $B$. Near the pole of the FR $B_{\rm pole}$ we approximate this dependence as linear and obtain the Zeeman energies $E^{(Z)}_j = - \mu_j (B-B_{\rm pole})$ for $j\in\{a,g,e\}$ with the magnetic moments $\mu_j$. We choose an interaction picture that makes $E^{(Z)}_a$ vanish and obtain $\Delta_g = \mu_{ag} (B-B_{\rm pole})/\hbar$ and $\Delta_e = - \Delta_L + \mu_{ae} (B-B_{\rm pole})/\hbar$ with $\mu_{ag}=2\mu_a-\mu_g$ and $\mu_{ae}=2\mu_a-\mu_e$. The bound-to-bound transition is treated in an interaction picture with a rotating wave approximation, where $\Delta_L=\omega_L-\omega_{eg}$ is the detuning of $\omega_L$ from the bound-to-bound resonance $\omega_{eg}$ at $B=B_{\rm pole}$.

The parameter $\alpha$ with $|\alpha|^2 = U_{\rm bg} \mu_{ag} \Delta B/2\hbar^2$ describes the FR coupling between states $|a\rangle$ and $|g\rangle$, where $\Delta B$ is the width of the FR. The parameter $\beta\propto E_0$ describes PA from state $|a\rangle$ to $|e\rangle$, which turns out to be negligible for the parameters of our experiment.

We assume that states $|g\rangle$ and $|e\rangle$ never obtain a large population because $\alpha\psi_a$ and $\beta\psi_a$ are both small compered to $\gamma_e$. Hence, we can adiabatically eliminate states $|g\rangle$ and $|e\rangle$ and obtain a complex-valued scattering length in analogy to Ref.\ \cite{abeelen:99}
\begin{eqnarray}
\label{a}
\lefteqn{
a = a_{\rm bg} + \frac{m}{2\pi\hbar} 
} \nonumber \\ && \times
\frac{ |\alpha|^2 (\Delta_e-i\gamma_e/2) + |\beta|^2 \Delta_g - {\rm Re}(\alpha^* \Omega_R^* \beta)}{|\Omega_R/2|^2- \Delta_g (\Delta_e-i\gamma_e/2)}
.
\end{eqnarray}
The two-body loss rate coefficient for a BEC is \cite{duerr:09} $K_2=-8\pi\hbar {\rm Im}(a)/m$. The FR used in our experiment has the following parameters \cite{volz:03,duerr:04,duerr:04a}: $B_{\rm pole}=1007.4$ G, $a_{\rm bg}=100.5 a_0$, $\mu_a/2\pi\hbar=1.02$ MHz/G, $\mu_{ag}/2\pi\hbar=3.8$ MHz/G, and $\Delta B=0.21$ G. A fit to the data in Fig.\ \ref{fig-Autler-Townes} yields the following best-fit values for the parameters of the bound-to-bound transition $\omega_{eg}/2\pi=382,046,942.8\pm0.3$ MHz, $|d_{eg}|/ea_0=0.28\pm 0.05$, $\gamma_e/2\pi=4.7\pm0.5$ MHz, $\mu_{ae}/2\pi\hbar=2.6\pm0.1$ MHz/G, and a negligible result for $\beta$. Here $a_0=52.92$ pm is the Bohr radius and $e=1.602\times10^{-19}$ C the elementary charge. The lines in Fig.\ \ref{fig-avoided-crossing}{\bf b} are plotted for $|d_{eg}|/ea_0=0.31$. We attribute the deviation from the other data to a technical problem with the light intensity calibration. Note that our model assumes that $\beta/E_0$ is constant. A possible FR enhancement of the PA resonance \cite{mackie:08,junker:08} is not considered here.

\section*{Acknowledgments}
We thank B. Bernhardt and K. Predehl for providing light from their frequency comb. We acknowledge fruitful discussions with J.~I.\ Cirac and T.\ Volz. This work was supported by the German Excellence Initiative via the Nanosystems Initiative Munich and by the Deutsche Forschungsgemeinschaft via SFB 631.

\end{document}